%% file: ms.tex
\documentclass[12pt,preprint]{aastex}

\begin{document}

\title{
A Deep, High-Resolution Survey at 74 MHz
}

\author{A.~S.~Cohen \altaffilmark{1,2},
H.~J.~A.~R\"ottgering \altaffilmark{3}, 
M.~J.~Jarvis \altaffilmark{3,4},
N.~E.~Kassim \altaffilmark{1},
T.~J.~W.~Lazio \altaffilmark{1}
}

\altaffiltext{1}{US Naval Research Laboratory, Remote Sensing Division, Code 7213, Washington, DC, 20375; aaron.cohen@nrl.navy.mil, namir.kassim@nrl.navy.mil} 
\altaffiltext{2}{National Research Council Postdoctoral Fellow}
\altaffiltext{3}{Leiden University, Sterrewacht, Oort Gebouw, P.O. Box 9513, 
2300 RA Leiden, The Netherlands, rottgeri@strw.LeidenUniv.nl}
\altaffiltext{4}{Astrophysics, Department of Physics, Keble Road, Oxford, OX1
3RH, UK, mjj@astro.ox.ac.uk}

\begin{abstract}

We present a 74 MHz survey of a 165 square degree region located near 
the north galactic pole.  This survey has an unprecedented combination of 
both resolution ($25''$ FWHM) and sensitivity ($\sigma$ as low as 
24 mJy/beam).  We detect 949 sources at the $5\sigma$ level in this region, 
enough to begin exploring the nature of the 74 MHz source population.  We 
present differential source counts, spectral index measurements and the size 
distribution as determined from counterparts in the high resolution 
FIRST 1.4 GHz survey.  We find a trend of steeper spectral indices for the 
brighter sources.  Further, there is a clear correlation between spectral 
index and median source size, with the flat spectrum sources being much 
smaller on average.  Ultra-steep spectrum objects 
($\alpha \leq -1.2$; $S_{\nu} \propto S^{\nu}$) are identified, and we 
present high resolution VLA follow-up observations of these sources which, 
identified at such a low frequency, are excellent candidates for high 
redshift radio galaxies.  

\end{abstract}

\section{Introduction}

The new 74 MHz system on the Very Large Array (VLA), fully implemented in 
1998 \citep{1993AJ....106.2218K}, has opened a new window into the 
previously unexplored regime of very low frequency radio observations at 
high sensitivity and sub-arcminute resolution.  

The radio source population at very low frequencies is of interest for 
several reasons.  Samples selected at $\nu \sim 74$~MHz are completely
dominated by isotropic radio emission, unlike those found at higher
frequencies, where orientation-dependent Doppler boosting enhances
the observed emission for some fraction of sources: even for the 
178~MHz 4C survey (\citet{1965MmRAS..69..183P}; \citet{1967MmRAS..71...49G}) 
10\% of sources were significantly affected by Doppler beaming 
\citep{Wall97}, leading to biased samples of radio sources where those 
sources with jets pointing toward us are relatively over represented.  In 
addition, at frequencies below 100~MHz spectral curvature is much more 
common than at higher frequencies, providing an important tool for studying 
the properties of the absorbing gas.  This is because at low frequencies, 
the needed optical depths for absorption due to free-free absorption by 
H\,{\sc ii} regions (intrinsic or intervening) or synchrotron self-absorption 
require lower electron densities than at higher frequencies. 

The ability to find steep spectrum sources is of value because of its
relative efficiency in finding very high redshift radio galaxies ($z >
2$).  Identifying ultra-steep spectrum sources has now been used by
many groups for a number of years to find the highest redshift radio
galaxies \citep{rott94, 1996ApJS..106..215C, blundell98}.  This method
has been implemented mainly with frequencies above 150 MHz 
(\cite{2000A&AS..143..303D}, \cite{blundell98}), 
and is based on the curved nature of the radio spectral energy
distribution.  Most low-redshift radio galaxies have spectra that 
flatten below 325 MHz, while the higher redshift
objects are observed at a higher rest frame frequency, where the SED 
is steep.  At 74 MHz,
this method is likely to be even more efficient, since the spectra of
most low-redshift radio galaxies will have flattened to a much greater
degree at this very low frequency, increasing the contrast between
them and the high redshift radio galaxies.

Although radio galaxies are no longer the only galaxies found at the
very highest redshifts \citep{Hu02, Kodaira03, Cuby03} and certainly
not the most abundant high-redshift galaxies
\citep{1996ApJ...462L..17S, 1999ApJ...519....1S} they do have unique
characteristics: (1) the selection via radio emission means that
radio-loud AGN are selected free of caveats concerning dust, which is
probably a very important factor in the high-redshift Universe, as is
evidenced by the huge sub-millimeter luminosities associated with
radio galaxies at high redshift (e.g. Archibald et al. 2001; Reuland
et al. 2003) (2) their strong narrow-emission lines allow
relatively simple redshift determinations that do not rely on stellar
continuum breaks or absorption lines and (3) powerful 
($L_{151} > 10^{25}$~W~m$^{-2}$~Hz$^{-1}$) 
radio galaxies inhabit some of the most
massive galaxies in the Universe at all cosmic epochs
\citep{2001MNRAS.326.1585J, 2002AJ....123..637D, 2003MNRAS.339..173W},
and as such provide a unique probe into the build-up of massive
galaxies. With the recent work on the correlation between black-hole
mass and bulge luminosity in the local Universe \citep{1998AJ....115.2285M}, 
the fact that galaxies which exhibit powerful radio
emission are the most massive means they probably host the the most
massive black holes \citep{dun03, 2003A&A...399..869B}.
The evidence for super-massive black holes in the centers of all bulge
dominated galaxies also implies that every massive galaxy in the local
Universe may have had an active phase in the past. This is further
reinforced by the similarity in number density of super-massive black
holes in the local Universe and the density of quasars at $z \sim
2.5$, i.e. the quasar epoch. Thus, we may need to understand AGN
activity to understand massive galaxy formation in general.

A further characteristic of radio galaxies is, by definition, their
radio emission. This allows a unique estimate of the time since the
radio AGN was triggered \citep{1999AJ....117..677B}, and it enables us 
to probe the gaseous environment via Faraday rotation measures
\citep{1997ApJS..109....1C}.  But possibly more importantly in the
years to come, it allows us to probe 21~cm HI absorption along the
line-of-sight to the radio source. This will be vitally important as
21~cm absorption (and emission) will allow us to probe the epoch of
reionization. These studies
will be tractable with the development of the new era of radio
telescopes such as the LOFAR\footnote{http://www.lofar.org} 
\citep{2000SPIE.4015..328K} and the 
SKA\footnote{http://www.skatelescope.org} which will have the 
sensitivity to probe such a signal.  
Radio sources that are placed at the epoch before which the reionization
is fully completed will be a very important probe of this epoch, allowing
the study of the neutral gas at parsec and kiloparsec scales.  Such scales
are much smaller than are being probed by WMAP or Plank.  Thus, it is 
crucial to find radio sources at $z > 6$ to allow these investigations to 
take place, and selecting ultra-steep spectrum (USS) radio sources is a
well proven technique for finding the highest-redshift radio galaxies
\citep{1996Natur.383..502R, 1999ApJ...518L..61V}, and selecting at
74~MHz allows us to do this better than ever before.

In this paper, we describe the survey with a discussion of observational
methodology in Section 2 and the mapping results and source extraction in 
Section 3.  In Section 4 we explore the nature of the 74 MHz
radio source population.  In Section 5, we identify the ultra-steep spectrum
sources and present high-resolution follow-up radio observations to 
determine their identities.  

\section{Observational Methodology}\label{sec:methods}

\subsection{The Data Set}

The data used to produce this survey comes from observations taken
on March 7, 1998, intended to map two normal galaxies at 74 MHz
(NGC 4565 and NGC 4631).  
Though these two targets were largely resolved out at this resolution, the 
combination of A-configuration resolution, favorable ionospheric 
conditions and pointings near the north galactic pole where the 
background temperature is low produced the deepest observation
ever obtained below 100 MHz.  That the two fields observed were separated by 
$6.4^\circ$, which is roughly the radius of the primary beam at 74 MHz,
allowed the two fields to be ideally combined to produce a single 
deep image of an area roughly $17^\circ\times10^\circ$ in size.

The data was taken in spectral line mode to reduce bandwidth smearing
and to facilitate removal of radio frequency interference.  The 
total bandwidth of 1.54 MHz was divided into 128 channels, which 
Hanning smoothing reduced to 64.  The central frequency was 73.8 MHz
and the integration time was 10 seconds.
Each of the two pointings was observed for 3 hours, and the observation
alternated between the two fields to better distribute the $uv$-coverage.

\subsection{Challenges for High Resolution Imaging at Low Frequency}

There are many challenges to high resolution imaging at low radio
frequencies which have prevented this observational regime from being
explored until the past few years.  In this section, we summarize these
issues and their solutions.  \citet{2003K} provide a more comprehensive 
description of 74 MHz observations with the VLA.  In addition, a 
low frequency data reduction tutorial is available online
\footnote{http://rsd-www.nrl.navy.mil/7213/lazio/tutorial}.

\subsubsection{Radio Frequency Interference}

The monitor and control system of the VLA generates significant radio 
frequency interference (RFI) below 100 MHz, and there is the potential 
for external RFI as well.  For this reason (and others) the data are 
acquired in spectral-line mode.  Excision of potential RFI is performed 
on a per-baseline basis for each visibility sample.  Actual excision
is accomplished by correcting the spectral-line data for the shape of
the VLA bandpass at this frequency, fitting a linear baseline, then
flagging any data exceeding a level typically set to be 6--8 times the
expected rms noise level.\footnote{This algorithm is contained within 
the \textsc{aips} task \texttt{FLGIT}.}  Typically this procedure 
removes 10-15\% of the visibility data.  It is notable, however, that 
at 74~MHz there is no evidence to indicate that any of this RFI is 
externally generated. 

\subsubsection{The ``3-D Problem''}

One crucial aspect for the post processing of these data sets is the
need for a three-dimensional inversion of the visibility data.  
A conventional two-dimensional inversion of the three-dimensional
visibility function, as measured by a non-coplanar (i.e.,
non--east-west) array like the VLA, introduces errors in the
image plane which increase as the square of the distance from the 
phase center.  The
``3-D'' problem becomes severe at long wavelengths where the primary
beam is large and contains hundreds of discrete sources which must be
properly deconvolved to achieve thermal-noise-- or 
classical-confusion--limited maps.

To solve this problem, \cite{cp92} have developed a polyhedron 
algorithm in which the three-dimensional ``image volume'' is approximated 
by many two dimensional ``facets,'' small enough that the traditional 
two-dimensional assumption is valid.  Thus a ``fly's eye'' of small 
overlapping images, each with its own phase center, is made to cover the 
large field of view.  Typically, many hundreds of facets are needed to map 
the full primary beam area, and after cleaning and imaging, these images 
are then combined into a single undistorted image of the full field of 
view.

\subsubsection{Ionospheric Image Distortion}
\label{fb.sec}
The biggest challenge to achieving high resolution at low frequencies
is the distortion of phases caused by the ionosphere, an effect which 
increases proportionally with wavelength \citep{1993AJ....106.2218K}.  
This problem is compounded by the fact that the field of view at 74 MHz 
is so large, that these
phase distortions can vary significantly across the field of view.
This puts phase calibration of 74 MHz data into an entirely different 
regime from higher frequencies in which all phase distortions can be 
completely described by a single time-variable number for each antenna.
For this reason, traditional self-calibration does not apply.  Rather,
a field-based calibration method, in which the phase calibration is 
position-dependent, is needed.  Such an algorithm has been written 
by J. J. Condon and W. D. Cotton\footnote{\texttt{VLAFM}: a 
special-purpose task designed to work within \textsc{aips}} 
\citep{2002URSI....C}.  This algorithm models the ionosphere as phase 
screen varying in space and time, 
and applies a two-dimensional Zernike polynomial phase correction to 
correct the phases across the entire field of view.

\subsection{Data Reduction}

Cygnus A (3C405) was used for bandpass and amplitude calibration.  
Phase calibration was more complicated.  As described in section \ref{fb.sec}, 
calibration which applies a constant phase correction to the entire field 
is insufficient for imaging of the entire 74 MHz VLA primary beam, and 
field-based calibration is needed.  However, an initial calibration 
estimate is necessary to begin the field-based calibration, and so as a 
first step, we calibrated the field to a sky model obtained using the 
1.4 GHz NVSS catalog \citep{1998AJ....115.1693C} adjusted for a spectral 
index of $\alpha = -0.75$\footnote{\textsc{aips} task \texttt{FACES}}.  
Next, field-based calibration was used to remove ionospheric smearing 
before imaging and cleaning were performed\footnote{\texttt{VLAFM}: a 
special-purpose task designed to work within \textsc{aips}}.
The field-based calibration
was performed with a time resolution of 2 minutes, enough to significantly 
detect an average of 6 to 10 sources in the field of view.  Each pointing 
was mapped to a radius of $6^{\circ}$ using roughly 300 facets.  We used a 
pixel size of $7.5''$ and each facet was 360 pixels in diameter.  A circular 
$25''$ restoring beam was used.  The images from the two fields were then 
corrected for the primary beam shape, and co-added to produce a single image.

\section{Results}\label{sec:results}

\subsection{Image Quality}

\begin{figure}
\epsscale{0.80}
\plotone{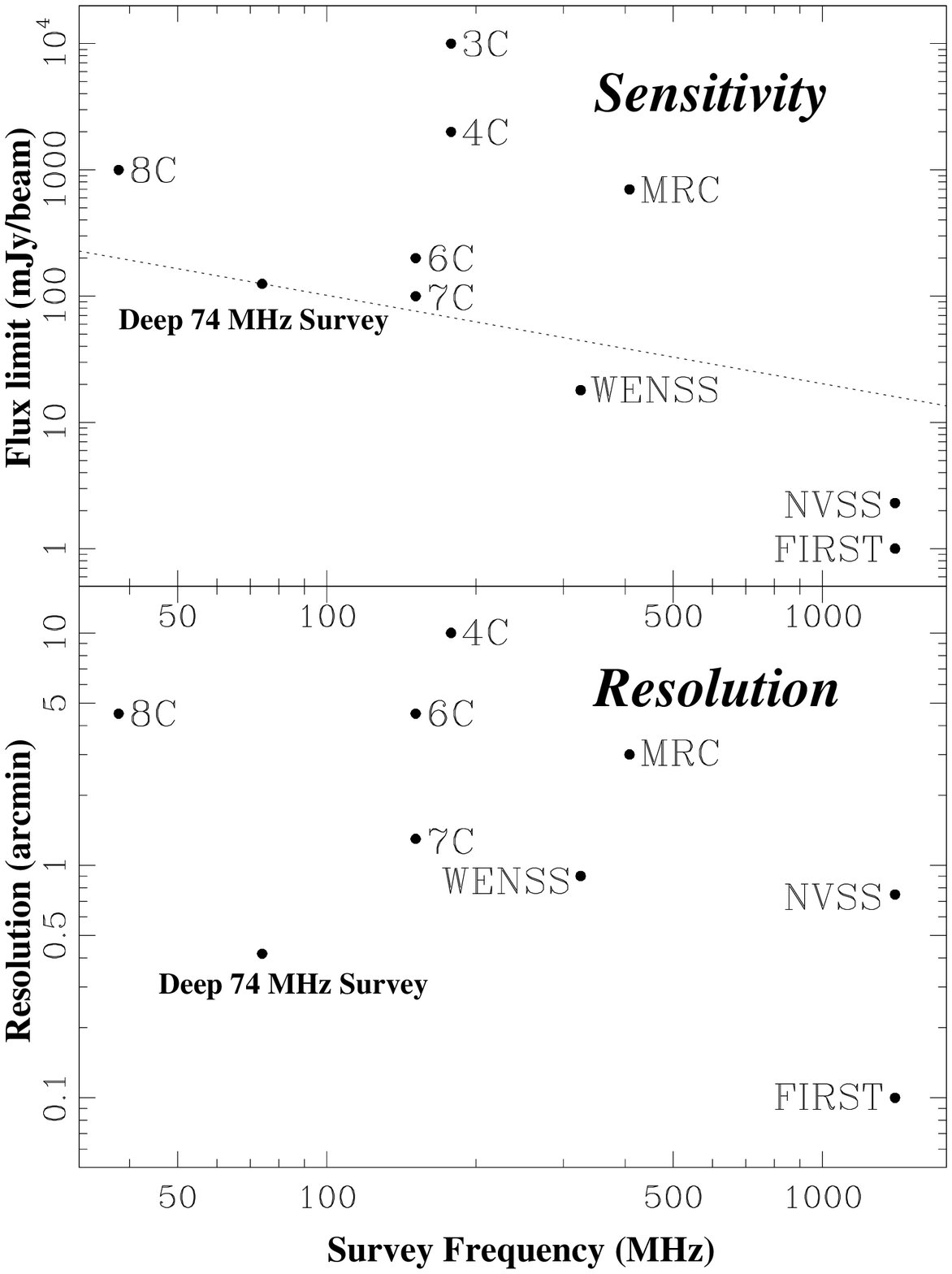}
\caption{The sensivitiy ($5\sigma$) and resolution of our deep 74 MHz
radio survey compared to that of the major low frequency radio surveys.
The dotted line on the sensitivity plot (top) shows the flux density of a 
source at our sensitivity limit assuming a spectral index of $\alpha = -0.7$. 
\label{survey.fig}}
\end{figure}

\begin{figure}
\epsscale{0.80}
\plotone{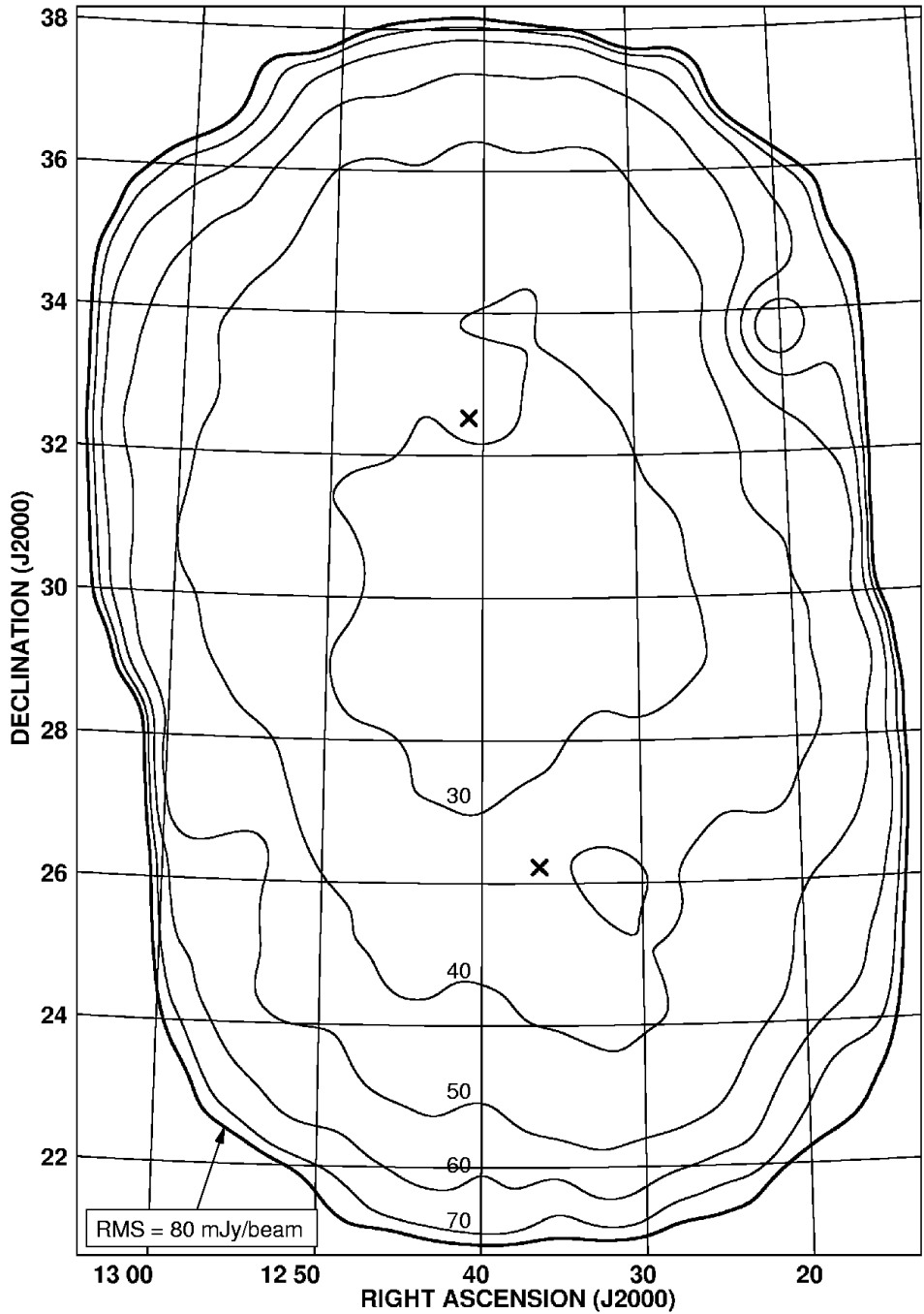}
\caption{Noise levels in the 74 MHz survey area as determined by local 
measuments of the RMS noise in the image.  The contours plotted
are at 30, 40, 50, 60, 70 and 80 mJy/beam, with the outermost contour
at 80 mJy/beam in bold defining the boundary of the survey.  Each of the
two field centers is marked with an ``X''.
\label{noise.fig}}
\end{figure}

Our final image has a resolution of $25''$ and a minimum RMS noise level of 
24 mJy/beam, a unique combination for such a low frequency survey 
(see Figure \ref{survey.fig}).  However, the noise level varies over the image
due to the primary beam sensitivity patterns for the two fields, the 
added sensitivity due to the two fields overlapping and occasional
incompletely cleaned sidelobes from bright sources (Figure \ref{noise.fig}).  
The image has no real ``edge'', just an increase in the noise level 
with distance from the pointing centers.  If we define the survey area as 
the area in our map for which the RMS noise is lower than a set limit, 
then as this limit is increased the survey area increases.  However, at points 
more than about $5^\circ$ from either pointing center the noise level 
begins to increase so rapidly that the benefits of increasing the 
survey area quickly diminish (see Table \ref{tableA}).  We chose the field 
edge to be at the 80 mJy/beam noise level, which defines a total survey area 
of 165.1 square degrees, with almost half of this area having a sensitivity 
below 40 mJy/beam.

\subsection{Source Extraction}

The same algorithm which was written for use in the 1.4 GHz 
NVSS \citep{1998AJ....115.1693C} was used to identify and 
characterize sources in this 74 MHz survey survey\footnote{\texttt{VSAD}: 
a special-purpose task designed to work within \textsc{aips}}.
This algorithm works by identifying ``islands'' of emission in the 
image above a set threshold and fitting each source to a model of up to four 
Gaussians.  We set as our source detection threshold that sources 
must have both a peak and integrated flux level at least 5 times the 
local RMS noise level.  Therefore, the source detection threshold
varies over the image as the RMS noise varies.  

\begin{figure}
\epsscale{0.80}
\plotone{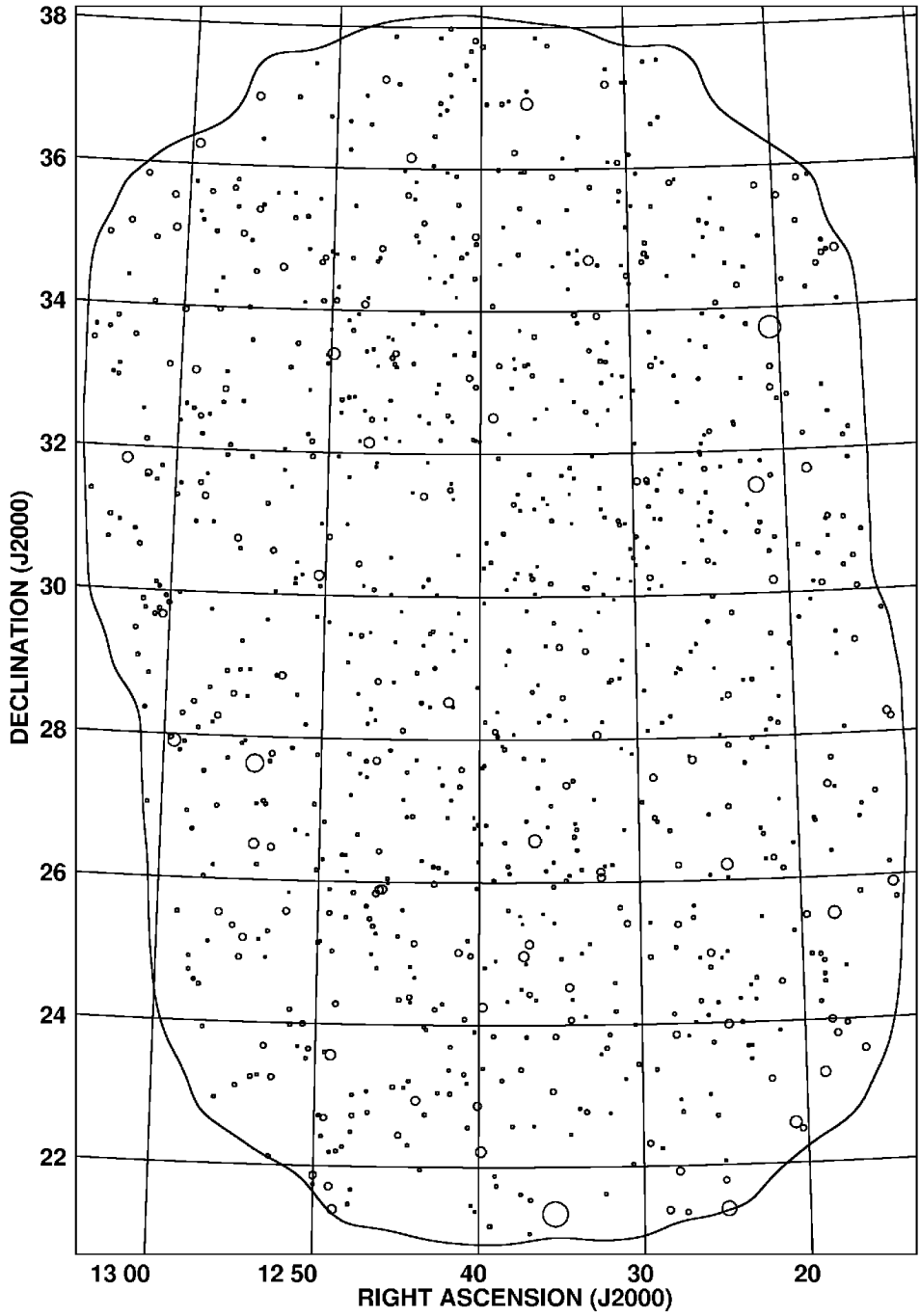}
\caption{Location of all sources identified at the $5\sigma$ level within 
the survey region.  The area of each circle is proportional to the 
integrated flux of that source.  The weakest source is 146 mJy and 
the strongest is 39.9 Jy.  
\label{sources.fig}}
\end{figure}

\begin{figure}
\epsscale{0.85}
\plotone{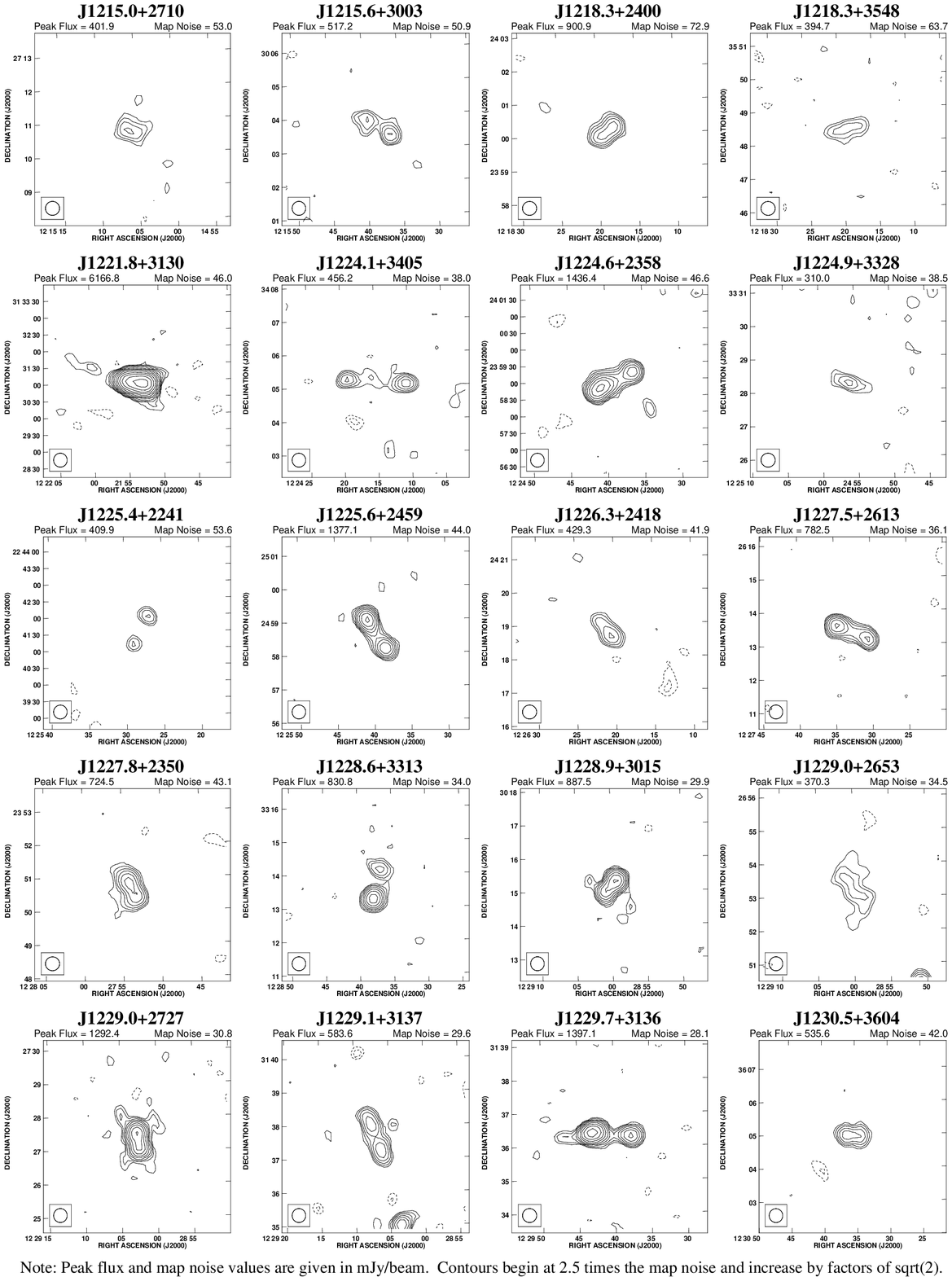}
\caption{ (page 1 of 4)
Maps of the 80 largest objects detected at 74 MHz.  These objects all have 
a fitted source size (for single sources) or maximum separation between 
components (for multiple sources) of at least $48.4''$. 
\label{large.fig}}
\end{figure}

We combined into multiple sources (sources composed of more than one 
Gaussian component), those sources located within $120''$
of each other.  This distance was chosen because it corresponds to 
the separation distance of radio sources for which doubles which are
components of the same source (usually a radio galaxy) outnumber those 
that are actually two separate objects by a ratio of 10 to 1 as
determined by a recent study of the clustering of radio sources in the 
NVSS \citep{2003A&A...405...53O}.
In total, we detected 949 sources (Figure \ref{sources.fig})
with integrated fluxes ranging from 146 mJy to 39.9 Jy.  
Of these, 70 were multiple sources.  Most sources were not much larger
than the $25''$ beamsize, with only 75 sources (7.9\%) measuring larger 
than two beamwidths across, either as the fitted size of a single source
or the maximum separation within a multiple source.  Most of the extended 
sources appear to be FRII type radio galaxies.  Figure \ref{large.fig}
shows contour maps of the 80 largest sources in the survey.  The complete 
catalog of all sources is available in the electronic edition of the 
{\it Astrophysical Journal} in Table 2.

\section{Source Statistics}

\subsection{Source Counts}

We present Euclidean-normalized differential source counts in Figure
\ref{lnls.fig}.  These were calculated within flux density bins of size
$\pm15\%$ from the central flux density.  To ensure a true measure of 
source densities, for each flux density bin we only considered the region
of the survey in which all sources in that flux density bin are detected 
with at least $7\sigma$ confidence, the canonical level of $2\sigma$ 
completeness for sources selected above a $5\sigma$ threshold.  Thus, 
for each flux density bin we only selected sources from the region of 
the survey with a noise level lower than 1/7 of the lower flux density 
limit of that bin.  The source density was calculated by dividing this 
source count by the area of that ``$7\sigma$'' region from which the 
sources were selected.

Also plotted in Figure \ref{lnls.fig} is a polynomial fit to the source 
count at 327 MHz \citep{1991PhDT.......241W} and its adjustments according 
to various spectral indices.  The data at 74 MHz seems to correspond to an 
average spectral index of somewhere between $\alpha = -0.5$ and 
$\alpha = -0.75$, though there seems to be an overall trend of
steeper corresponding spectral index at higher flux levels.  We 
caution that this is not a direct measurement of the average 
spectral index as many compact sources seen at 327 MHz are not seen at 
74 MHz due to synchrotron self-absorption.

We note that the differential source counts function depends to some degree 
on our choice of spatial cutoff for grouping sources.  Though we grouped 
components closer than $120''$ into single sources, if larger physical 
sources were broken into multiple ``sources'', the differential source 
counts function would be artificially increased for lower flux densities 
and decreased for the highest flux densities.  However, this effect is 
likely to be minor.  For example, if we double the spatial cutoff to 
$240''$, only 35 additional sources ($3.7\%$) are ``grouped''.  Further, 
it is unlikely that all of these new multiple sources are physical, with 
\citep{2003A&A...405...53O} predicting near parity between physical and 
coincidental doubles at a spacing of $240''$.

\subsection{Comparisons to Other Survey Catalogs}

We compared our source catalog to the 1.4 GHz NVSS catalog 
\citep{1998AJ....115.1693C} which, with its resolution of $45''$, is well 
enough matched to our resolution to provide accurate spectral index 
measurements.  
The long frequency interval between 74 MHz and 1.4 GHz also contributes to 
the accuracy of the spectral index measurements as a flux ratio error of 
10\% produces a spectral index error of only $\Delta\alpha = 0.032$.  
We found NVSS matches within $60''$ for 947 out of our total of 949 sources.  
The two sources without NVSS matches were J1230.6+3247 and J1253.6+2509,
implying very steep spectral index upper limits of $-1.47$ and $-1.78$ 
respectively.  It is therefore reasonable to wonder if these are real 
detections.  The peak flux density of J1230.6+3247 was measured to be 
188 mJy/beam, which is a $6.3\sigma$ detection given the local noise level of 
29.8 mJy/beam.  This relatively strong detection and the fact that the 
spectral index necessary to explain the lack of detection is, though 
very steep, not even the steepest in the survey, lead us to conclude that 
this source is probably real.  The peak flux density of J1253.6+2509 was 
measured to be 256 mJy/beam, which was just barely over our $5\sigma$ 
detection threshold for the local noise level of 50.4 mJy/beam.  Because
of this and the extraordinary steepness of its spectrum necessary to explain
its lack of detection by NVSS, we conclude that it is a non-negligible
possibility that this may be a false detection.  However, as it meets our 
criteria for a source detection, we still include it in our source list.

We present the spectral index measurements in Figure \ref{spec.fig}
plotted versus the integrated flux density at 74 MHz.  The spectral 
indices of all sources with respect to their NVSS counterparts
are included in the source catalog, available electronically.  We find
a median spectral index of $\alpha = -0.79$.  Figure \ref{spec.mean.fig}
shows the spectral index as a function of 74 MHz flux density, and it is 
clear that the average spectral index steepens with increasing flux 
density.

In addition to the spectral index between the widely spaced frequencies
of 74 MHz and 1.4 GHz, it is of interest to investigate the actual shape 
of the spectrum within this interval.  For this, we turn to the 327 MHz 
WENSS survey \citep{1997A&AS..124..259R}.  Although the WENSS only 
covers the northern sky down to a declination of $+29^{\circ}$, which 
falls near the middle of our survey region, we find counterparts 
(within $60''$) to 545 sources.  Figure \ref{color.fig} shows a comparison 
of the spectral indices for two frequency intervals in the form of a radio
color-color plot.  As this plot shows, most source spectra flatten 
considerably from the ``low'' frequency interval from 327 MHz to 1.4 GHz 
to the ``very low'' frequency interval from 74 MHz to 327 MHz.  
Quantitatively, we found a median change in the spectral index of 
$\Delta\alpha = 0.24$.

We also compared our source catalog to the higher resolution ($5.4''$)
1.4 GHz FIRST survey \citep{1997ApJ...475..479W}.  We found 
matches within $30''$ for 945 out of our total of 949 sources (including
neither of the 2 sources without NVSS matches).  This allowed a much better 
measurement of source sizes than was possible with the $25''$ resolution of
the 74 MHz observations.  We define the source size as either the deconvolved
major axis for single sources in FIRST or as the maximum separation
between components for the multiple sources.  Overall, we find a median
source size of $8.1''$, though we notice a strong correlation 
between source size and spectral index.  Figure \ref{spec.hist.fig}
shows the median source size as a function of spectral index along 
with a spectral index histogram.  It is 
clear that the steeper the spectrum the larger the sources, which makes
sense as one would expect compact sources to have flatter spectra on 
average due to synchrotron self-absorption, although this trend seems to 
stall if not reverse for the very steep spectrum objects ($\alpha < -1$).

\section{Ultra Steep Spectrum Sources}

\subsection{Identification and Follow-up}

To identify our sample of ultra-steep spectrum (USS) sources, we 
chose a spectral index cutoff of $\alpha_{74}^{325} < -1.2$.
With this criterion, we identified 26 USS sources out of a total of 
949 sources (2.7\%).  This included 24 sources with measured spectral 
indices and the 2 sources with no FIRST or NVSS counterparts 
which have spectral index upper limits which qualify them as USS as well.  

This percentage of USS sources is much higher than that found in the 
previous USS searche by \citet{{2000A&AS..143..303D}} which 
found a USS fraction of only 0.5\%.  Even if we use their
stricter USS criteria of $\alpha < -1.3$, we still have detected 18 USS 
sources, or 1.9\%.  
It is unclear exactly what causes this difference, but it is possibly due 
to a combination of effects.  First, by selecting sources at 74 MHz, rather 
than 325, we have eliminated virtually the entire population of flat 
spectrum objects from 
our sample.  Second, as 74 MHz observations are less sensitive than 
those at higher frequencies, we have a brighter sample, and it is known
that as the flux limit increases, the USS fraction increases.  

There are four main 
possibilities for the identities of such USS sources: (1) pulsars, (2) 
high redshift radio galaxies (HzRGs: $z > 2$), (3) cluster halos or relics 
(CHRs) and (4) fossil radio galaxies (FRGs: radio galaxies for which the 
energized particles have undergone significant spectral aging).  As this 
survey is located near 
the north galactic pole, we expect very few if any pulsars.  Using the 
most recent estimate for the radio luminosity function of CHRs 
\citep{2002A&A...396...83E}, we predict that on the order of 1 CHR will
be detectable in this region given our sensitivity limit, so this is a 
real possibility.  HzRGs, cluster 
halos and fossil radio galaxies can be distinguished by their morphologies.  
We expect small ($\leq30''$) FRII morphologies for HzRGs, while cluster halos 
or fossil radio galaxies will be much larger and more diffuse.  
It is difficult to qualify the morphology with the $25''$ resolution of our
74 MHz images, so again we turn to the 1.4 GHz FIRST survey 
\citep{1997ApJ...475..479W} with its resolution of $5.4''$.  In Figure 
\ref{first.fig} we present FIRST images of each of the 24 USS sources with 
an NVSS counterpart.  Comparison with the POSS-II shows matches or 
possible matches to 4 out of 24 sources, roughly consistent with the 15\%
reported by \citet{2000A&AS..143..303D}.  We have also begun our own 
follow-up observations 
at 1.4 GHz with the VLA in A-configuration which provides roughly $1.4''$ 
resolution.  So far we have so observed 18 of the 24 USS sources with NVSS 
counterparts and 11 of these were detected (Figure \ref{Aconf.fig}), with 
the remaining 7 likely resolved out.  A list of all USS sources
we found is provided in the print version of Table \ref{table2}, and 
brief descriptions of each USS source follows below.

\subsection{Brief Description of Individual USS Sources}

The following are brief notes on each USS source.  Source images at 1.4 GHz 
are described from FIRST ($5.4''$ resolution) and or VLA A-configuration 
follow-up ($1.4''$ resolution).  We speculate on the identity of each source 
from the four main classes of USS objects:  pulsar, cluster halo/relic (CHR), 
fossil radio galaxy (FRG) or high redshift radio galaxy (HzRG).

\noindent
{\bf J1225.0$+$2146:} The FIRST map shows a bright double lobed FRII 
with a separation of roughly $25''$.  
This is likely to be a HzRG or FRG.

\noindent
{\bf J1226.3$+$2418:} The FIRST map shows a $20''$ FRII with diffuse 
extension.  A-configuration resolved out this source.  It was identified
in the POSS-II so it probably isn't high redshift.  
This is likely to be a FRG.

\noindent
{\bf J1228.9$+$3114:} The FIRST map shows a diffuse blob of about $15''$
which is resolved out in A-configuration.
This is likely to be a high-$z$ CHR, HzRG or FRG.

\noindent
{\bf J1229.1$+$3040:} Unresolved in FIRST but a fit to the A-configuration
map shows a source size of about $1''$.  This is likely a HzRG.

\noindent
{\bf J1229.9$+$3712:} Head-tail morphology in FIRST map.  
This is likely to be a HzRG or FRG.

\noindent
{\bf J1230.2$+$2326:} Faint double source in FIRST maps, resolved out
in A-configuration.  It was identified in the POSS-II so it probably isn't 
high redshift.  This is likely to be a FRG.

\noindent
{\bf J1230.6$+$3247:} No FIRST or NVSS counterpart. 

\noindent
{\bf J1231.2$+$2538:} A-configuration shows core-jet morphology with an
extent of $12''$.  
This is likely to be a HzRG or FRG.

\noindent
{\bf J1231.3$+$3724:} Unresolved in FIRST.  
This is likely to be a pulsar or HzRG.

\noindent
{\bf J1231.5$+$3236:} Resolved out in FIRST and A-configuration.  
This is likely to be a high-$z$ CHR or FRG.

\noindent
{\bf J1232.2$+$2814:} Unresolved in FIRST and A-configuration maps.  
This is likely to be a pulsar or HzRG.

\noindent
{\bf J1232.6$+$3157:} A-configuration shows a $7''$ FRII with much 
diffuse emission.  
This is likely to be a high-$z$ CHR, HzRG or FRG.

\noindent
{\bf J1234.3$+$2605:} Unresolved in FIRST but a fit to the A-configuration
map shows a source size of about $1''$.  This is likely a HzRG.

\noindent
{\bf J1238.2$+$2613:} Unresolved in FIRST but a fit to the A-configuration
map shows a source size of about $0.5''$.  This is likely a HzRG.

\noindent
{\bf J1238.8$+$3559:} FIRST and A-configuration show a well defined 
$17''$ FRII.  
This is likely to be a HzRG or FRG.

\noindent
{\bf J1243.7$+$2830:} FIRST and A-configuration show a faint $2''$
elongated object.  
This is likely to be a HzRG or FRG.

\noindent
{\bf J1245.9$+$3320:} FIRST shows a roughly $15''$ FRI morphology, which is 
resolved out in A-configuration.  
This is likely to be a high-$z$ CHR or FRG.

\noindent
{\bf J1246.4$+$2516:} FIRST shows a roughly $10''$ blob, which is 
resolved out in A-configuration.  
This is likely to be a high-$z$ CHR or FRG.

\noindent
{\bf J1248.2$+$2747:} FIRST and A-configuration show a faint $6''$
FRII morphology.  
This is likely to be a HzRG or FRG.

\noindent
{\bf J1249.0$+$3615:} FIRST shows a roughly $10''$ FRI morphology.  
This is likely to be a high-$z$ CHR or FRG.

\noindent
{\bf J1249.7$+$3408:} Unresolved in FIRST but a fit to the A-configuration
map shows a source size of about $0.8''$.  This is likely a HzRG.

\noindent
{\bf J1250.4$+$2941:} FIRST shows a $20''$ FRI morphology, which is 
mostly resolved out in A-configuration.  
This is likely to be a HzRG or FRG.

\noindent
{\bf J1252.7$+$2207:} FIRST shows a $20''$ diffuse blob.  
This is likely to be a high-$z$ CHR, HzRG or FRG.

\noindent
{\bf J1253.4$+$2703:} FIRST shows a very diffuse object of about $40''$
diameter which is resolved out in A-configuration.  
This is an ideal CHR candidate, though no nearby clusters are known.

\noindent
{\bf J1253.6$+$2509:} No FIRST or NVSS counterpart. 

\noindent
{\bf J1256.9$+$2811:} FIRST shows a $15''$ FRI morphology.  
This is likely to be a high-$z$ CHR, HzRG or FRG.

\section{Conclusions and Future Plans}

We have completed a deep, high resolution survey at 74 MHz and identified
949 sources.  We measured the differential source counts between roughly 
0.22 and 11 Jy, and find them consistent with those found at 327 MHz 
\citep{1991PhDT.......241W} adjusted by a spectral index of roughly 
$-0.75 < \alpha < -0.5$.  Comparison with the 1.4 GHz surveys NVSS and 
FIRST shows that the average source in our survey is steeper than this.
This could be explained if a significant fraction of sources become flat or 
inverted between 327 MHz and 74 MHz, and so are selected against by our 
flux limited survey.  We find that fainter sources tend to have slightly 
flatter spectra.  Source size studies with the FIRST survey indicate that
the flat spectrum objects are much smaller than the steeper spectrum objects
on average.  

A small fraction of sources, 26 out of 949, qualify as USS sources with 
spectral indices steeper than $\alpha < -1.2$.  We present a list of these 
sources, along with FIRST images of 24 of these and higher resolution 
A-configuration VLA images of 18 USS sources.  We find that all but a few
of these are good candidates for high redshift radio galaxies, while the
others are probably fossil radio galaxies or cluster halo, relic systems.
We plan optical and spectroscopic follow-up on the HzRG candidates to 
measure their redshifts.

\begin{center}
Acknowledgements \\
\end{center}

We thank W. C. Erickson for providing us with the data used for this 
survey.  The authors made use of the database CATS 
\citep{1997adass...6..322V} of the Special Astrophysical Observatory.  
This research has made use of the NASA/IPAC Extragalactic Database (NED)
which is operated by the Jet Propulsion Laboratory, Caltech, under contract
with the national aeronautics and spece administration.  The Second Palomar 
Observatory Sky Survey (POSS-II) was made by the California Institute of 
Technology with funds from the National Science Foundation, the National 
Geographic Society, the Sloan Foundation, the Samuel Oschin Foundation, and 
the Eastman Kodak Corporation.  ASC acknowledges fellowship support from 
the National Reseach Council.  Basic research 
in radio astronomy at the NRL is supported by the Office of Naval Research.

\newpage

\addtocounter{figure}{-1}
\begin{figure}
\epsscale{0.85}
\plotone{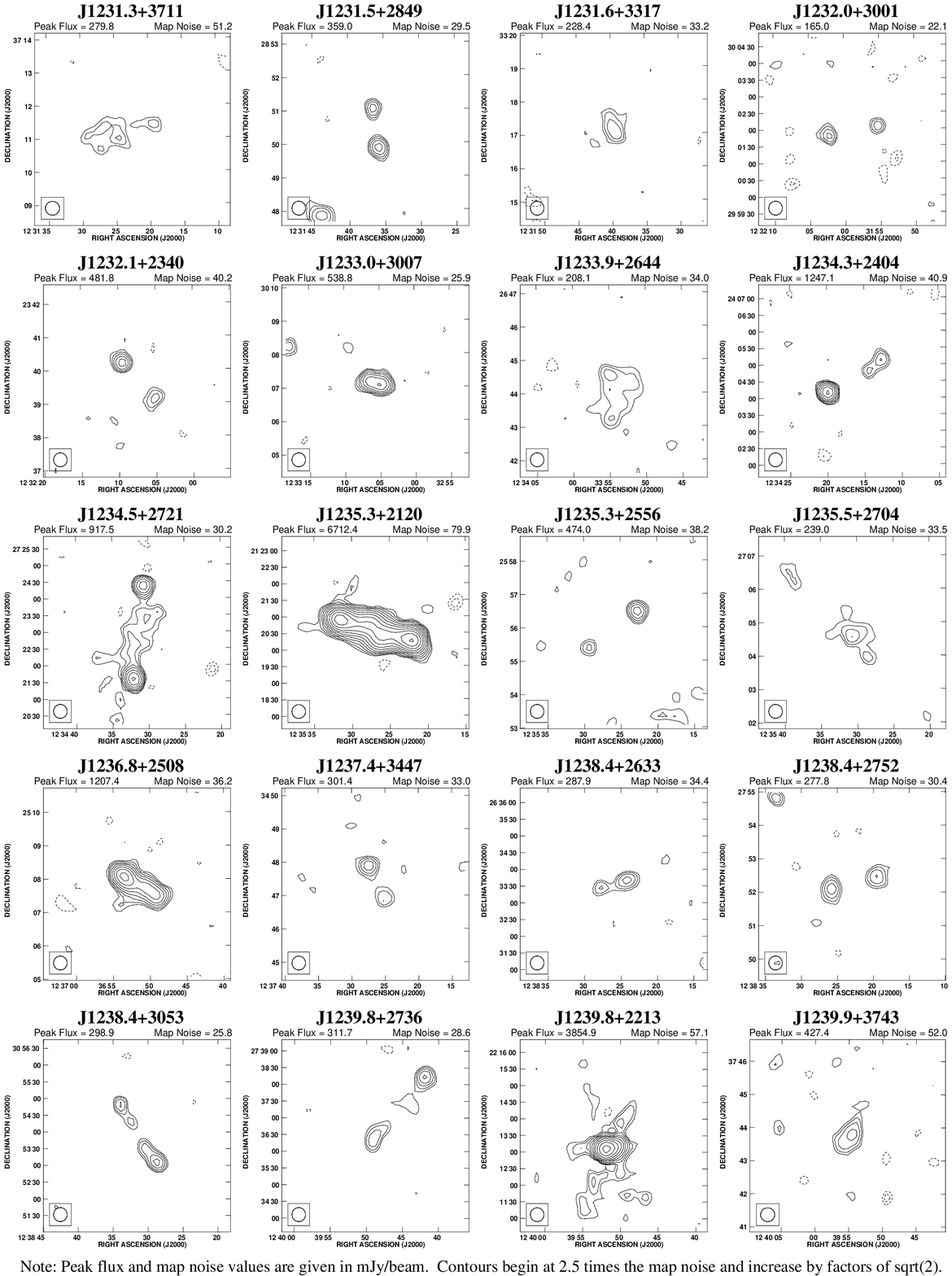}
\caption{ (page 2 of 4)
Maps of the 80 largest objects detected at 74 MHz.  These objects all have 
a fitted source size (for single sources) or maximum separation between 
components (for multiple sources) of at least $48.4''$. 
}
\end{figure}

\addtocounter{figure}{-1}
\begin{figure}
\epsscale{0.85}
\plotone{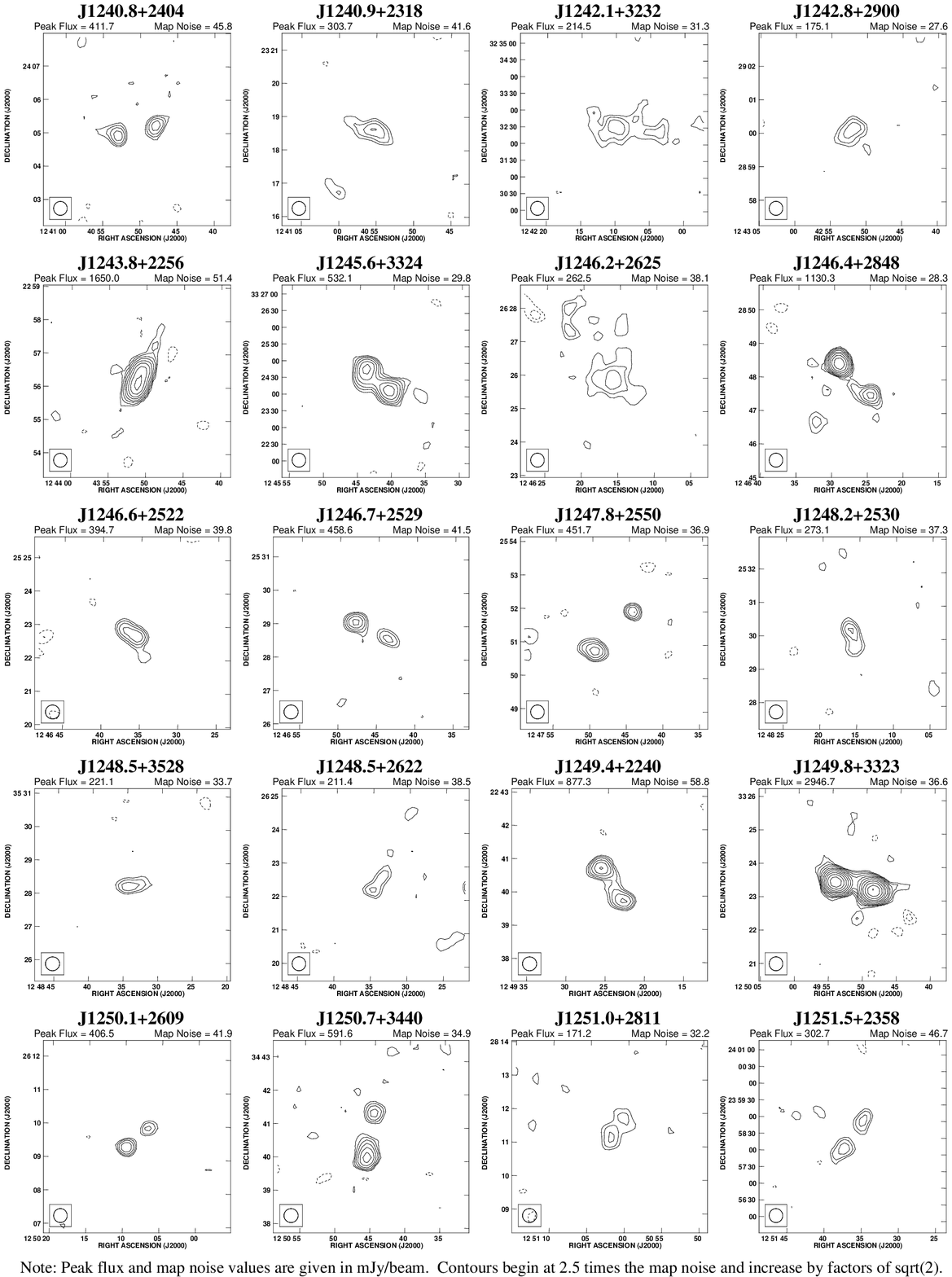}
\caption{ (page 3 of 4)
Maps of the 80 largest objects detected at 74 MHz.  These objects all have 
a fitted source size (for single sources) or maximum separation between 
components (for multiple sources) of at least $48.4''$. 
}

\end{figure}
\addtocounter{figure}{-1}
\begin{figure}
\epsscale{0.85}
\plotone{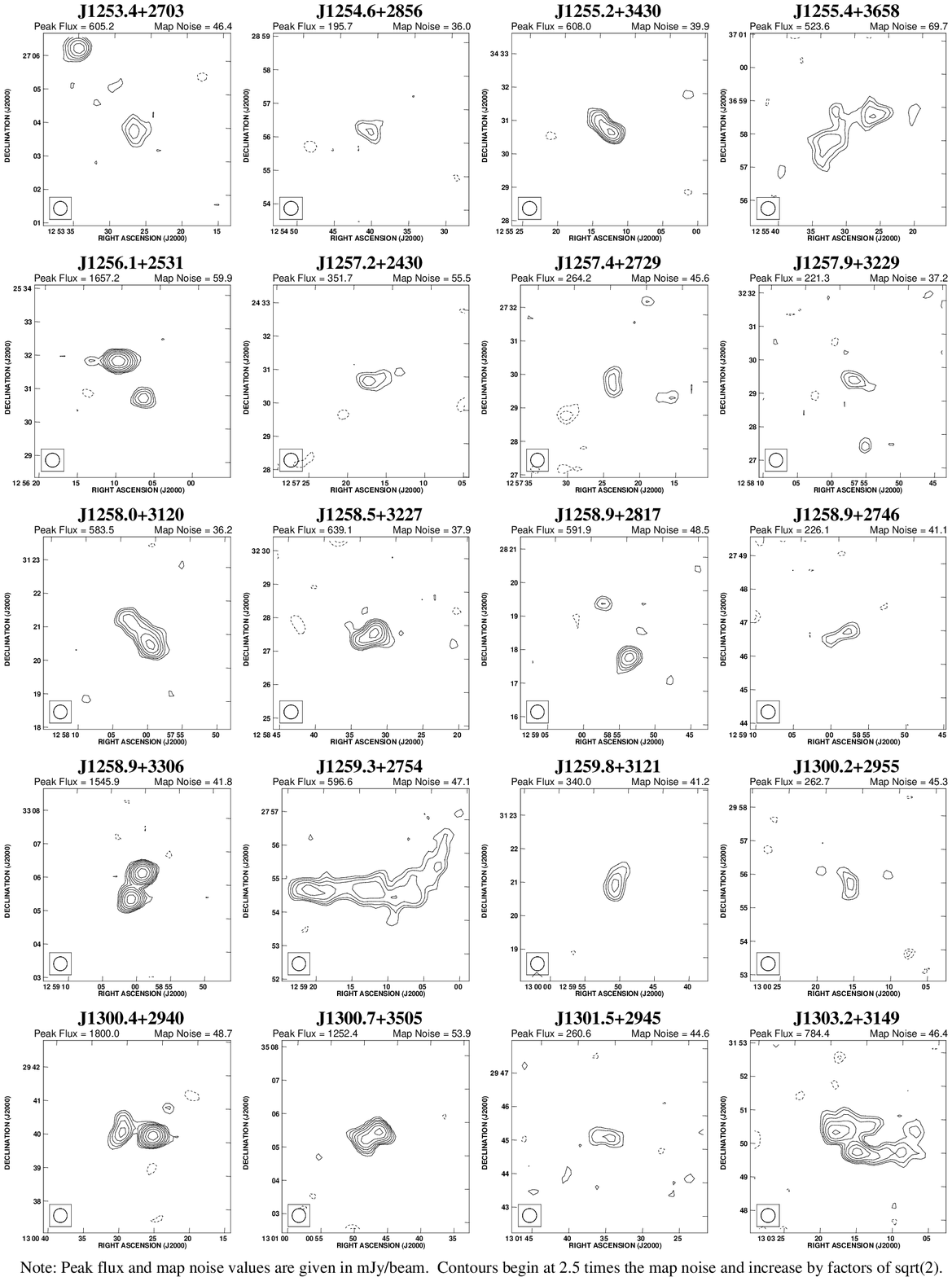}
\caption{ (page 4 of 4)
Maps of the 80 largest objects detected at 74 MHz.  These objects all have 
a fitted source size (for single sources) or maximum separation between 
components (for multiple sources) of at least $48.4''$. 
}
\end{figure}

\begin{figure}
\epsscale{0.80}
\plotone{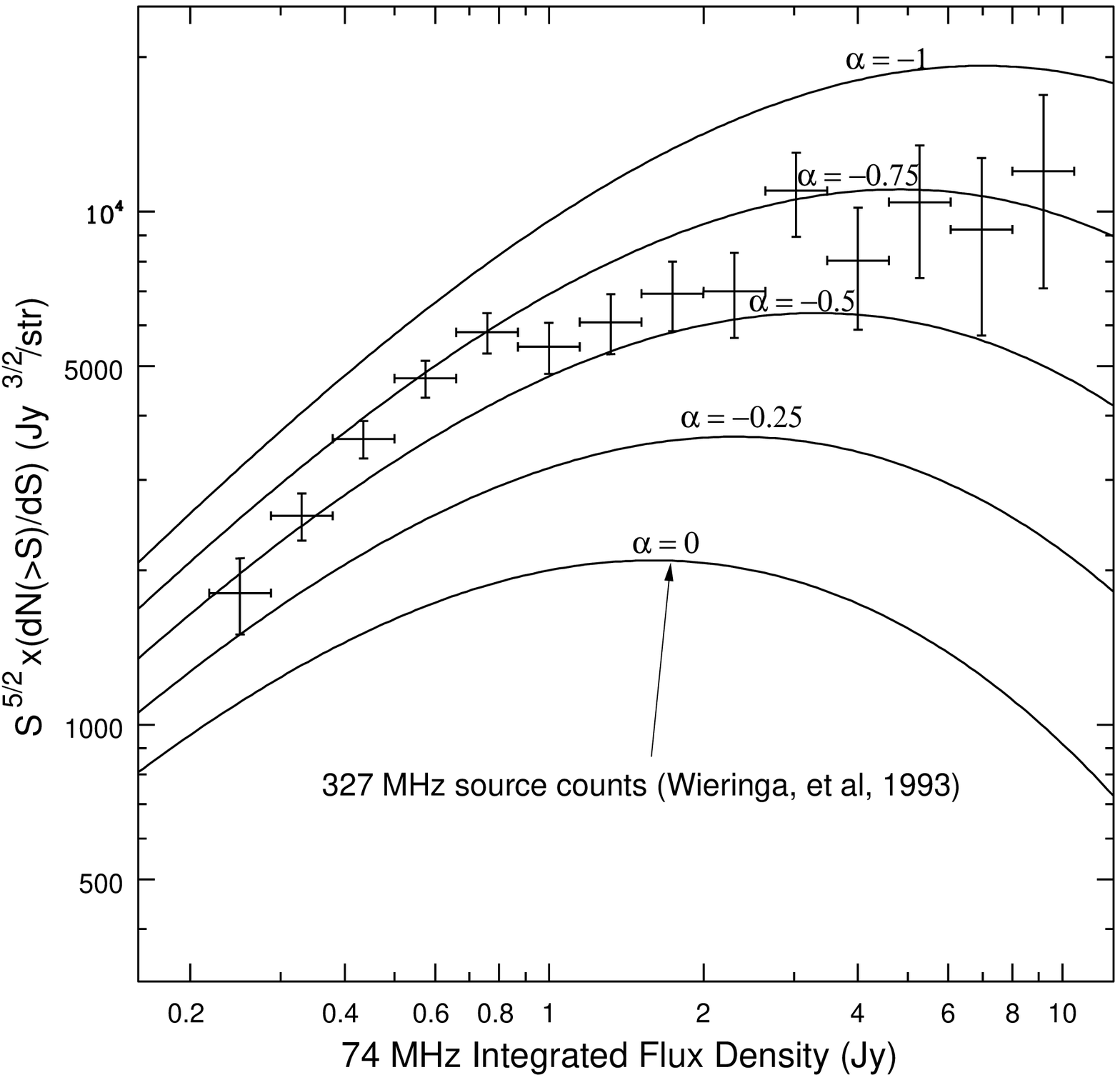}
\caption{
Euclidean-normalized differential source count.  Also plotted are 
source counts at 327 MHz \citep{1991PhDT.......241W} adjusted for various 
spectral indices.
\label{lnls.fig}}
\end{figure}

\begin{figure}
\epsscale{0.95}
\plotone{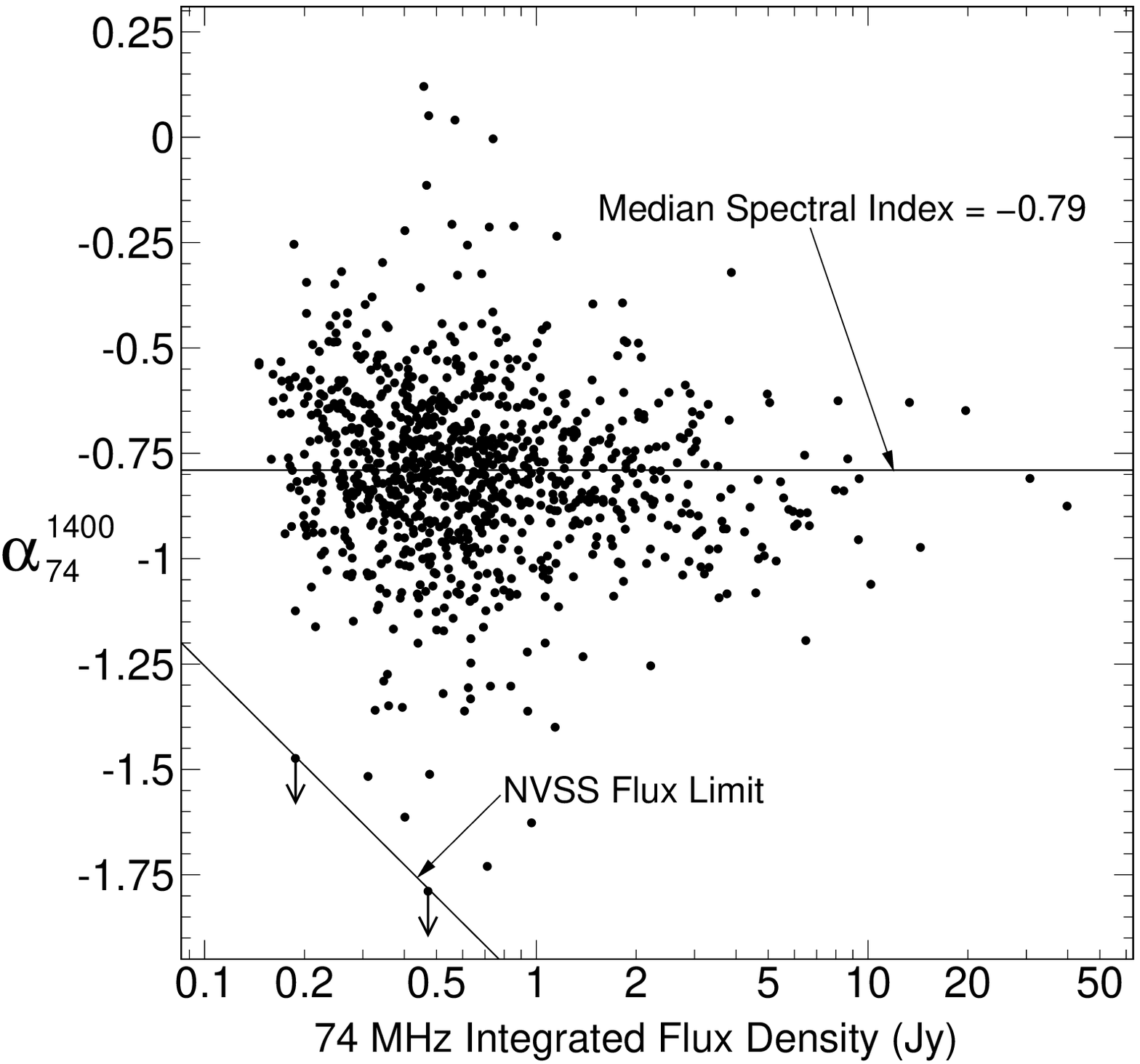}
\caption{
Spectral indices ($\alpha_{74}^{1400}$) versus 74 MHz integrated flux 
density.  Spectral index upper limits are plotted for the two sources 
without NVSS detections.
\label{spec.fig}}
\end{figure}

\begin{figure}
\epsscale{0.95}
\plotone{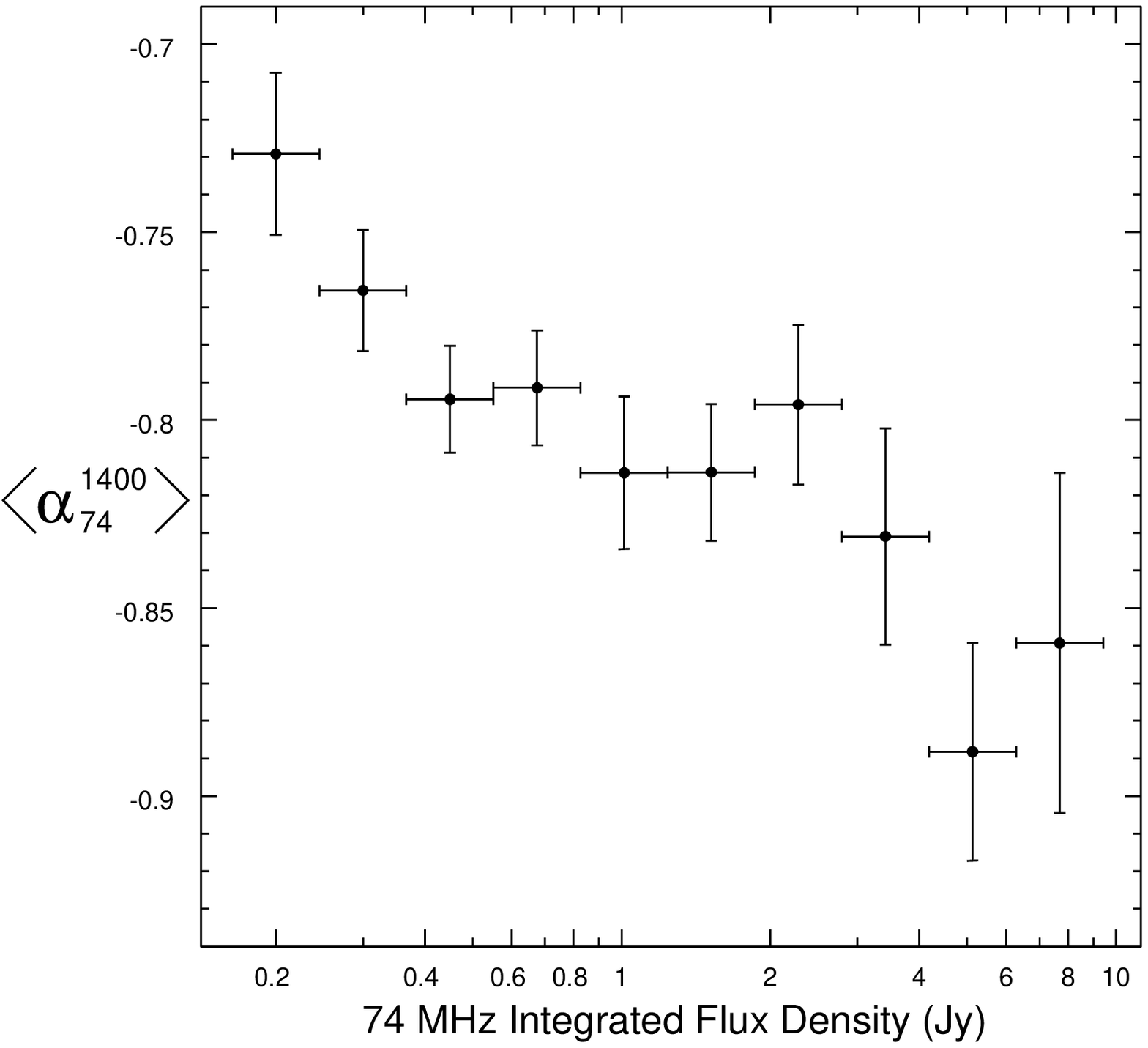}
\caption{
Mean spectral index versus 74 MHz flux density.  The horizontal
error bars correspond to the bin widths.
\label{spec.mean.fig}}
\end{figure}

\begin{figure}
\epsscale{0.95}
\plotone{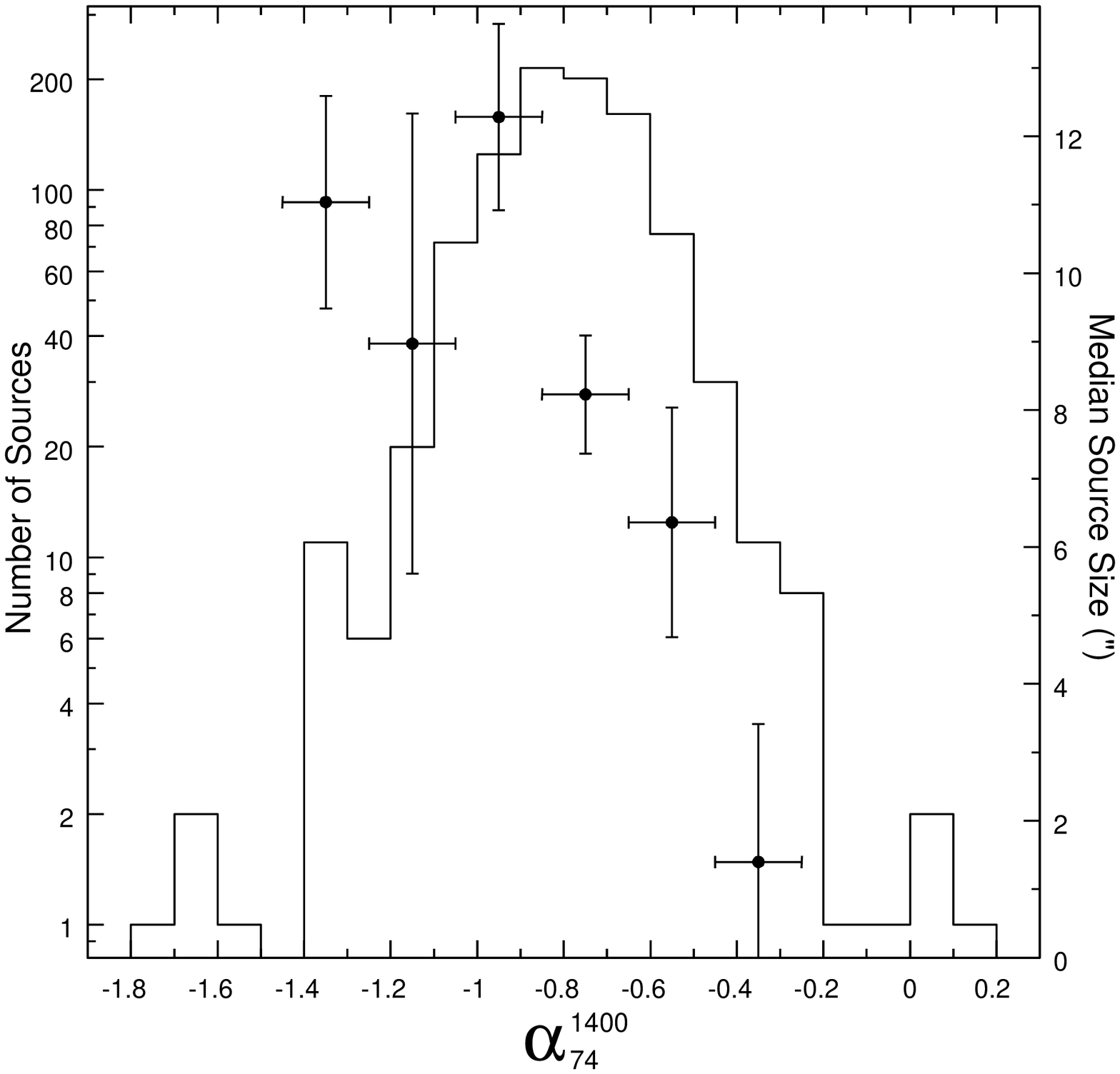}
\caption{
Spectral histogram along with median sources sizes as a function of 
spectral index.  Source sizes are determined with the FIRST images,
while spectral indices are determined using the 74 MHz integrated 
flux density and the NVSS flux densities.  Not enough sources were 
detected with $\alpha_{74}^{1400} < -1.45$ or $\alpha_{74}^{1400} > -0.25$
for meaningful statistics in those regions.
\label{spec.hist.fig}}
\end{figure}

\begin{figure}
\epsscale{0.95}
\plotone{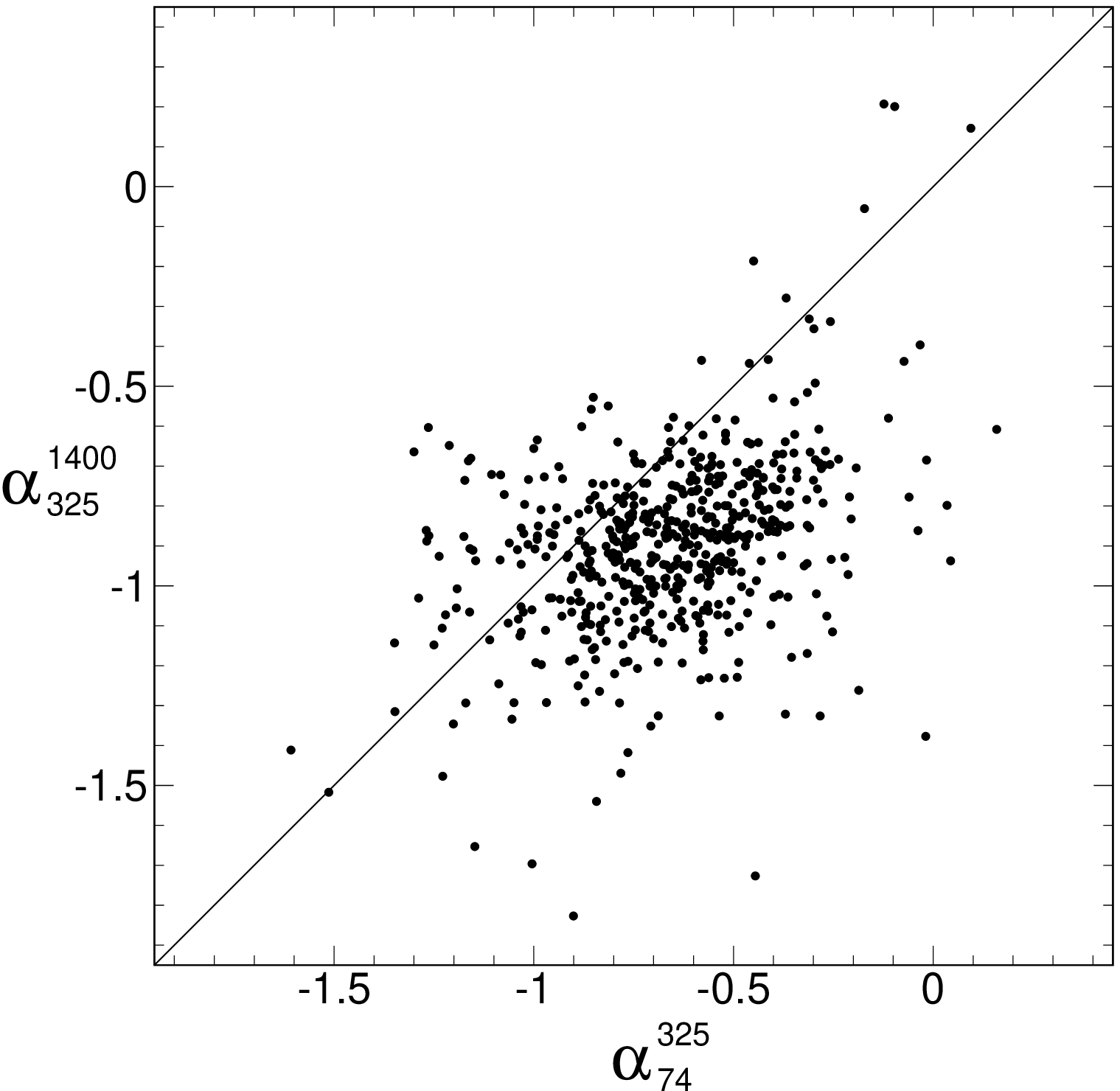}
\caption{
Radio color-color diagram of the 545 out of 949 sources with both an NVSS and 
a WENSS counterpart.  The diagonal line represents the location of sources
for which $\alpha_{74}^{325} = \alpha_{325}^{1400}$.  For most sources, 
the spectum flattens considerably in the lower frequency interval.
\label{color.fig}}
\end{figure}

\begin{figure}
\epsscale{1.0}
\plotone{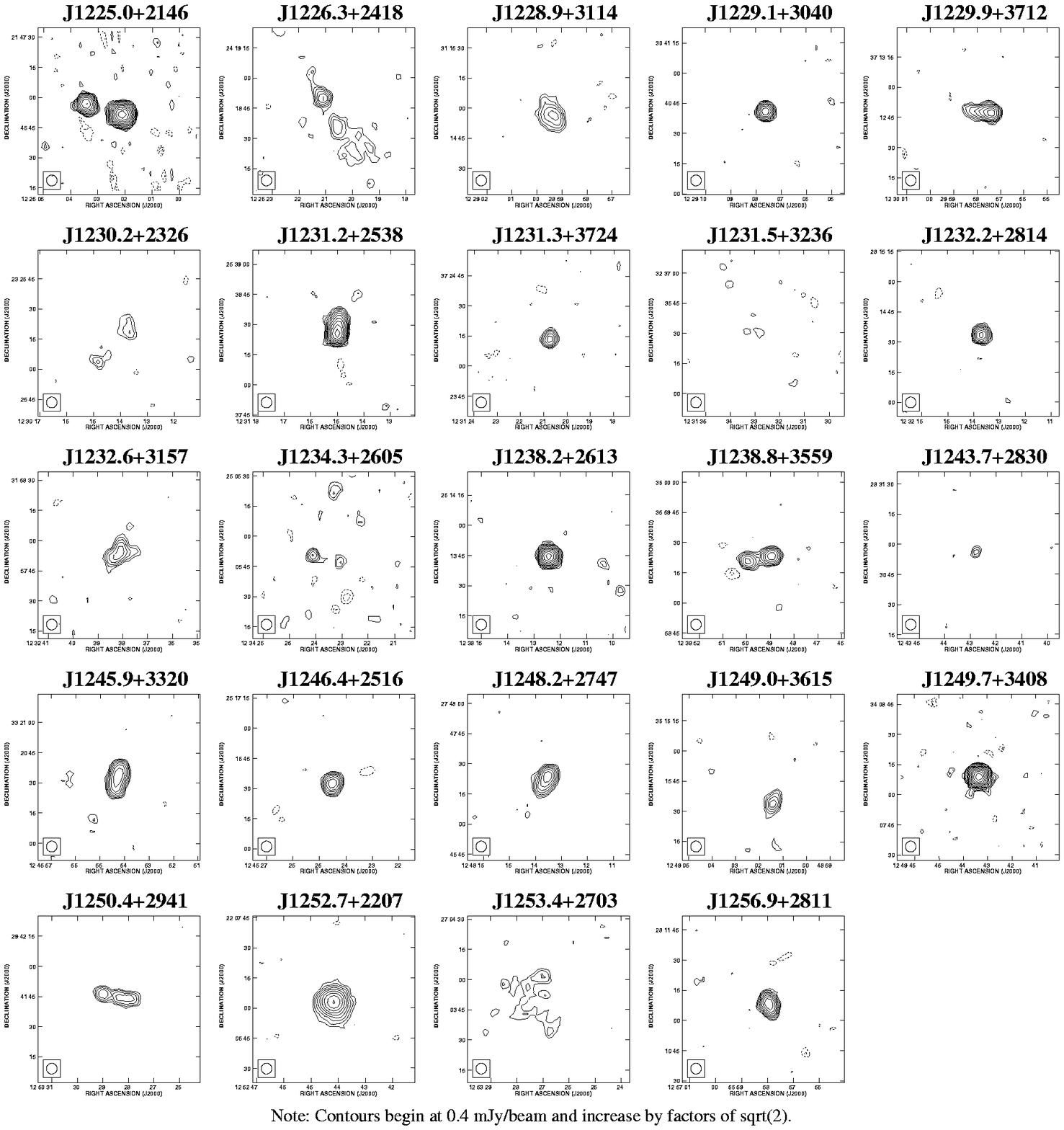}
\caption{
Images from the FIRST survey of the 24 out of 26 ultra-steep spectrum sources 
identified at 74 MHz, which had NVSS counterparts.  
\label{first.fig}}
\end{figure}

\begin{figure}
\epsscale{0.95}
\plotone{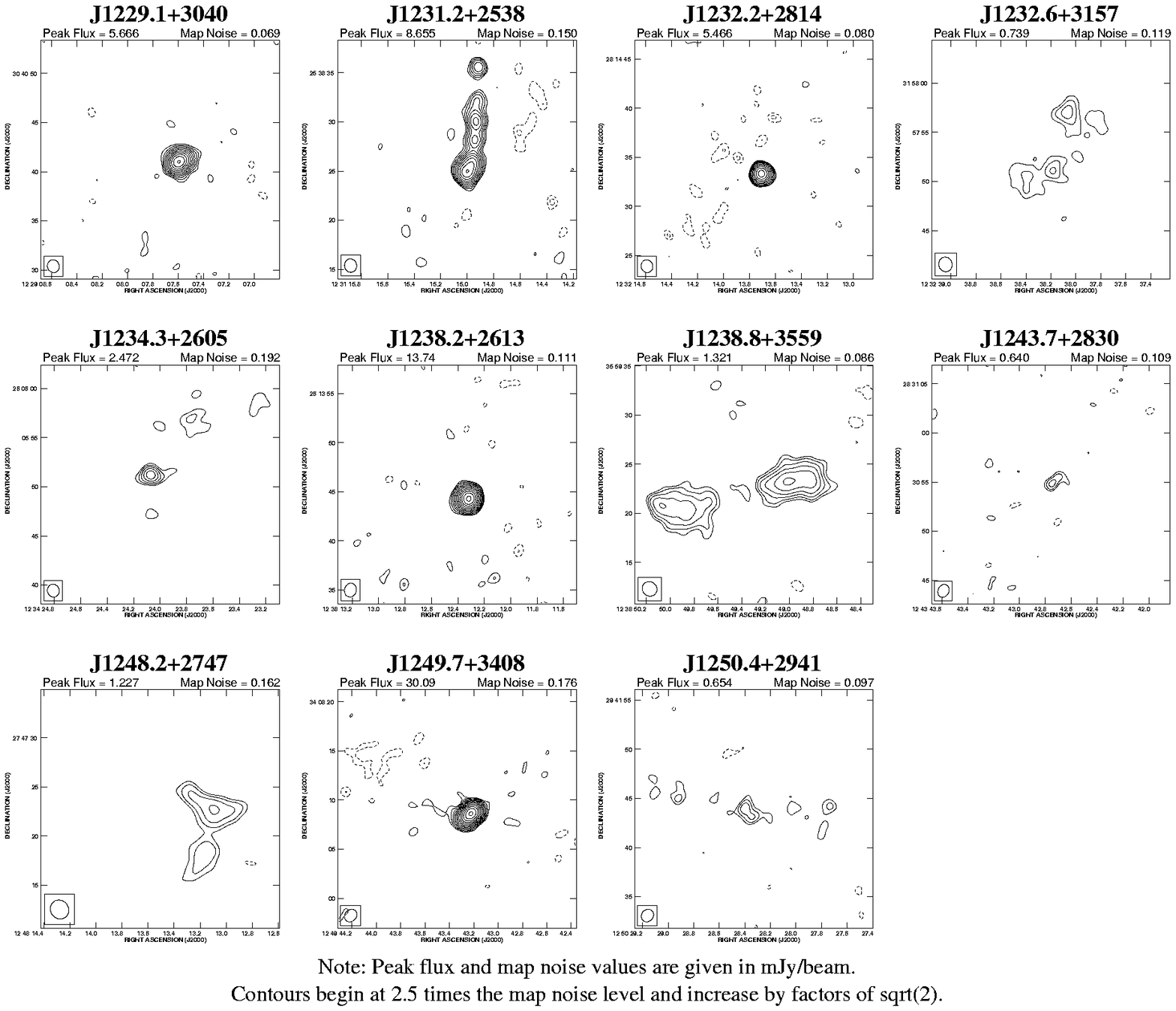}
\caption{
Images of the 11 detections from the 18 USS sources observed with the 
VLA A-configuration at 1.4 GHz.  The remaining 7 are diffuse 
enough to be resolved out at this resolution of roughly $1.4''$.  
\label{Aconf.fig}}
\end{figure}

\newpage 

\input{table1.tex}

\newpage 

\input{table2.tex}

\end{document}

%% file: table1.tex
\begin{deluxetable}{crr} 
\tablewidth{220pt}
\tabletypesize{\footnotesize}
\tablecaption{Survey Area \label{tableA}}  

\tablehead{
\colhead{Noise Limit} &
\colhead{Area} &
\colhead{Source Count}
\\
\colhead{(mJy/beam)} &
\colhead{(deg.$^2$)} &
\colhead{(number)}
}

\startdata

25 &   2.5 &  23 \\
27 &   8.6 &  62 \\
30 &  22.4 & 158 \\
35 &  50.0 & 356 \\
40 &  78.2 & 535 \\
50 & 119.1 & 765 \\
60 & 142.7 & 873 \\
70 & 156.9 & 924 \\
80 & 165.1 & 949 \\
\enddata
\end{deluxetable}

%% file: table2.tex
\begin{deluxetable}{lrrrrc} 
\tablewidth{360pt}
\tabletypesize{\footnotesize}
\tablecaption{USS Source List\label{table2}}  

\tablehead{
\colhead{Source} &
\colhead{$\alpha$ (J2000)} &
\colhead{$\delta$ (J2000)} &
\colhead{$S_{74,int}$} &
\colhead{$\alpha_{74}^{1400}$} &
\colhead{POSS-II ID} 
\\
\colhead{} &
\colhead{(h m s)} &
\colhead{($^{\circ}$ $'$ $''$)} &
\colhead{Jy} &
\colhead{} 
}
\startdata
J1225.0$+$2146 & 12 25 02.26 &  $+$21 46 52.5 & 2.213 & -1.25 & N \\
J1226.3$+$2418 & 12 26 20.65 &  $+$24 18 43.6 & 0.943 & -1.36 & Y \\
J1228.9$+$3114 & 12 28 59.58 &  $+$31 14 57.6 & 0.608 & -1.36 & N \\
J1229.1$+$3040 & 12 29 07.75 &  $+$30 40 41.2 & 0.356 & -1.27 & N \\
J1229.9$+$3712 & 12 29 58.03 &  $+$37 12 48.8 & 0.635 & -1.25 & N \\
J1230.2$+$2326 & 12 30 14.05 &  $+$23 26 17.2 & 0.968 & -1.63 & Y \\
J1230.6$+$3247 & 12 30 37.90 &  $+$32 47 21.9 & 0.188 & $< -1.47$ & \\
J1231.2$+$2538 & 12 31 15.08 &  $+$25 38 26.3 & 1.064 & -1.20 & ? \\
J1231.3$+$3724 & 12 31 20.77 &  $+$37 24 17.4 & 0.477 & -1.51 & N \\
J1231.5$+$3236 & 12 31 32.71 &  $+$32 36 30.1 & 0.311 & -1.52 & N \\
J1232.2$+$2814 & 12 32 13.76 &  $+$28 14 34.9 & 0.327 & -1.36 & N \\
J1232.6$+$3157 & 12 32 38.23 &  $+$31 57 52.1 & 0.359 & -1.35 & N \\
J1234.3$+$2605 & 12 34 23.25 &  $+$26 05 50.5 & 0.402 & -1.61 & N \\
J1238.2$+$2613 & 12 38 12.40 &  $+$26 13 45.8 & 0.728 & -1.30 & N \\
J1238.8$+$3559 & 12 38 49.30 &  $+$35 59 22.9 & 0.634 & -1.33 & N \\
J1243.7$+$2830 & 12 43 42.75 &  $+$28 30 55.5 & 0.712 & -1.73 & N \\
J1245.9$+$3320 & 12 45 54.20 &  $+$33 20 32.1 & 1.140 & -1.40 & N \\
J1246.4$+$2516 & 12 46 24.49 &  $+$25 16 35.1 & 0.347 & -1.29 & N \\
J1248.2$+$2747 & 12 48 13.51 &  $+$27 47 22.6 & 0.624 & -1.31 & N \\
J1249.0$+$3615 & 12 49 01.82 &  $+$36 15 46.8 & 0.440 & -1.20 & ? \\
J1249.7$+$3408 & 12 49 43.34 &  $+$34 08 09.2 & 1.383 & -1.23 & N \\
J1250.4$+$2941 & 12 50 28.33 &  $+$29 41 43.8 & 0.395 & -1.35 & N \\
J1252.7$+$2207 & 12 52 44.13 &  $+$22 07 04.9 & 0.940 & -1.22 & N \\
J1253.4$+$2703 & 12 53 26.85 &  $+$27 03 50.0 & 0.838 & -1.30 & N \\
J1253.6$+$2509 & 12 53 39.24 &  $+$25 09 57.7 & 0.472 & $< -1.78$ & \\
J1256.9$+$2811 & 12 56 58.15 &  $+$28 11 09.7 & 0.524 & -1.32 & N \\

\enddata
\tablecomments{The complete catalog is published in its entirety in the 
electronic edition of the {\it Astrophysical Journal}.}
\end{deluxetable}